\documentclass[sigconf]{acmart}
\usepackage[utf8]{inputenc}
\usepackage{hyperref}
\usepackage{booktabs}

\usepackage{wasysym}
\usepackage{pifont}
\usepackage{xspace}
\renewcommand\footnotetextcopyrightpermission[1]{}

\newcommand{\Par}[1]{\smallskip\noindent{\textbf{ #1.}}}
\newcommand{\cmark}{$\CIRCLE$}%
\newcommand{\xmark}{$\Circle$}%
\newcommand{\tild}{$\LEFTcircle$}%
\newcommand{\opc}{OPC UA\xspace}
\newcommand{\numproducts}{22\xspace}
\newcommand{\numlibraries}{16\xspace}
\newcommand{\trust}{Trustlist\xspace}
\newcommand{\rclient}{Rogue Client\xspace}
\newcommand{\rserver}{Rogue Server\xspace}
\newcommand{\middlep}{Middleperson\xspace}

\usepackage{graphicx}
\usepackage{msc}
\usepackage{subfigure}

\title{Security Analysis of Vendor Implementations of the OPC UA Protocol for Industrial Control Systems}

\author{Alessandro Erba}
\affiliation{%
   \institution{CISPA Helmholtz Center for Information Security\\
and
Saarbr\"ucken Graduate School of Computer Science, Saarland University}
   \country{Germany}}
\email{alessandro.erba@cispa.de}

\author{Anne Müller}
\affiliation{%
   \institution{CISPA Helmholtz Center for Information Security\\
and
Saarbr\"ucken Graduate School of Computer Science, Saarland University}
   \country{Germany}}
\email{anne.mueller@cispa.de}

\author{Nils Ole Tippenhauer}
\affiliation{%
   \institution{CISPA Helmholtz Center for Information Security}
   \country{Germany}}
\email{tippenhauer@cispa.de}

\begin{document}
\begin{abstract}
The OPC UA protocol is an upcoming de-facto standard for building Industry 4.0 processes in Europe, and one of the few  industrial protocols that promises security features to prevent attackers from manipulating and damaging critical infrastructures. Despite the importance of the protocol, challenges in the adoption of OPC UA's security features by product vendors, libraries implementing the standard, and end-users were not investigated so far.

In this work, we systematically investigate 48 publicly available artifacts consisting of products and libraries for OPC UA and show that 38 out of the 48 artifacts have one (or more) security issues. In particular, we show that 7 OPC UA artifacts do not support the security features of the protocol at all. In addition, 31 artifacts that partially feature OPC UA security rely on incomplete libraries and come with misleading instructions. Consequently, relying on those products and libraries will result in vulnerable implementations of OPC UA security features. To verify our analysis, we design, implement, and demonstrate attacks in which the attacker can steal credentials exchanged between victims, eavesdrop on process information, manipulate the physical process through sensor values and actuator commands, and prevent the detection of anomalies.

\end{abstract}
\maketitle
\pagestyle{plain}

\section{Introduction}

The increasing interconnection of industrial components critically relies on the security of the communication protocol adopted for Machine to Machine (M2M) communication to prevent adversarial manipulation of the process, leading to significant monetary loss and physical damage~\cite{waidnerIoT}. The OPC Unified Architecture (\opc) protocol~\cite{opcfoundation} is often considered the de-facto standard for building Industry 4.0 processes and it is the reference communication protocol in RAMI4.0 (the reference model for Industry 4.0~\cite{rami42015} in European countries such as France, Italy and Germany~\cite{Trilateral}). In 2006, the OPC Foundation released the first OPC UA specification featuring security mechanisms such as authentication, authorization, integrity, and confidentiality. 
The German Federal Office for Information Security (BSI) released the `\opc Security Analysis'~\cite{bsi17opcua}, reporting that `No systematic errors could be detected' in the \opc standard.

Nevertheless, there seem to be significant challenges to set up secure \opc deployments in practice. 
In~\cite{dahlmanns2020easing}, the authors perform a scan for \opc servers reachable over the Internet and find that 92\% of the  deployments show issues with security configuration. Among those, 44\% of the servers are accessible without any authentication requirements. To mitigate that issue, the authors suggest (similar to~\cite{bsi17opcua}) to disable insecure security modes and policies, and enforce user authentication. 
The authors of~\cite{dahlmanns2020easing} suggest that their findings are due to the configuration complexity of \opc, but no root cause analysis was provided. Indeed, \opc requires a correct configuration to prevent attacks such as: i) attackers feeding wrong information to clients (\rserver attack); ii) eavesdrop, and change values which can directly alter the physical process (\rclient attack); iii) or both (\middlep attack). To prevent such attacks, \opc specifies a number of alternative \emph{securityModes} (i.e.,~configurations for application authentication and encryption)%
, \emph{securityPolicies} (i.e.,~algorithms used for asymmetric encryption)
and \emph{UserIdentityToken} (i.e.,~configurations for user authentication). The main attacks to mitigate are i) \rserver 
, ii) \rclient 
, and iii) \middlep attacks.

In this work, we are the first to systematically investigate the security of a range of \opc libraries and products. 
 Despite the popularity of OPC UA and its advanced security concept (compared to  other industrial protocols), we show that in practice many  \opc artifacts have missing support for security features of the protocol. Those limitations  make the security configuration  infeasible, or give a false sense of security (i.e.,~traffic encryption occurs with untrusted parties). Several of the artifacts that claim to feature \opc security rely on libraries that are incomplete, insecure and provide instructions that lead to insecure settings. While arguably such issues are implementation-specific, we  show that the \opc standard is insufficient, as it supports those insecure behaviors. Lastly, we propose several  practical countermeasures.

To assess the vulnerability of libraries, we create a framework that implements 
\rserver, \rclient, and \middlep attacks. For each library, we set up a client and/or server application using protocol options that can be expected to provide a secured \opc instance, and then attack that system. For cases where the attacks succeed, we further investigate the cause. We show that \middlep attacks are possible 
and allow to manipulate the process and prevent reliable control of the system. Surprisingly, we find that even recovery of plaintext credentials from intercepted encrypted communications is easily possible with our \rserver and \middlep attacks. This finding is in contrast to prior work~\cite{polge2019assessing}, where the authors perform a \middlep attack on \opc applications and conclude that it is impossible to recover user passwords in \opc systems (even with security mode `None').

\Par{Contributions} We summarize our contributions as follows
\begin{itemize}
    \item We systematically analyze 22 products and identify significant reoccurring (for 15 out of 22) issues with the availability \opc security features, or their setup instructions.
    \item We systematically analyze 16 libraries and for all of them we identify at least one reoccurring issue with the implementation of security features (either at client side, at server side or both).
    \item We demonstrate the feasibility of the identified attacks by implementing the attacks with custom proof-of-concept applications\footnote{The framework is available at  \href{https://github.com/scy-phy/OPC-UA-attacks-POC}{\underline{https://github.com/scy-phy/OPC-UA-attacks-POC}}}.
\end{itemize}

\Par{Disclosure} We discussed our findings with the OPC foundation, leading to improvements of the specifications. Our work was based on publicly available information extracted form user manuals, library documentation and the \opc standard. As the feasibility of our attacks depends on (past) end-user configuration process carried out in the target system and not on a specific software/hardware vulnerability, there was no disclosure to vendors required. Nevertheless, our findings already led to improvements of at least one artifact after public discussion of this work.

\Par{Organization} The remainder of this work is organized as follows. In Section~\ref{sec:background} we present the details of the \opc protocol. In Section~\ref{sec:methodology} we present the our research methodology together with system and attacker model. In Section~\ref{sec:implementation} we explain how we build the framework that we use to analyze the  libraries. In Section~\ref{sec:results} we report the results of our research. A discussion of countermeasures is presented in Section~\ref{sec:contermeasures}. Related work is presented in Section~\ref{sec:relatedwork} followed by conclusions in Section~\ref{sec:conclusion}.

\section{Background on OPC UA}
\label{sec:background}

\begin{figure}[htb]
\centering
 \subfigure[]{\includegraphics[width=.28\textwidth]{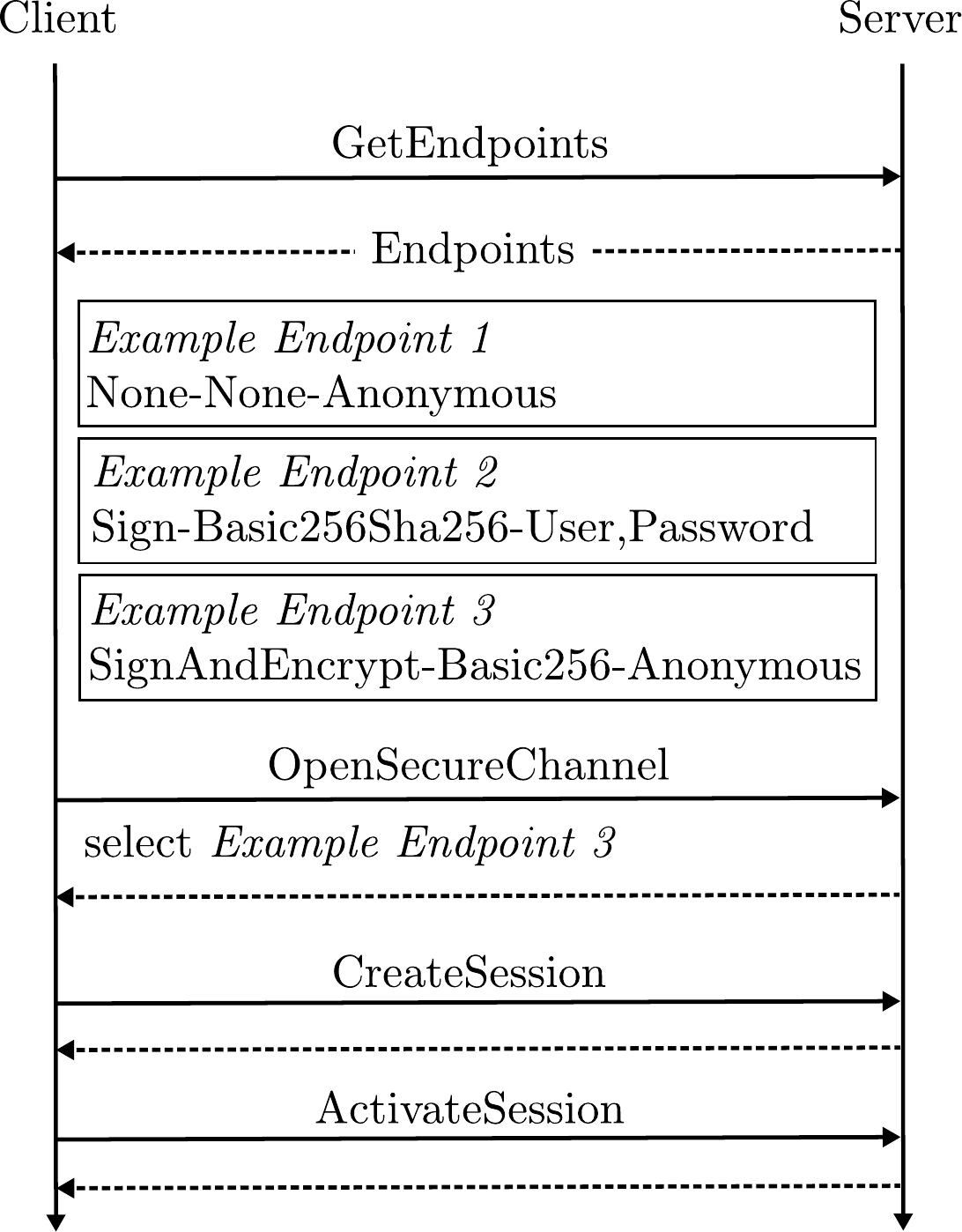}} 
 \subfigure[]{\includegraphics[width=.18\textwidth]{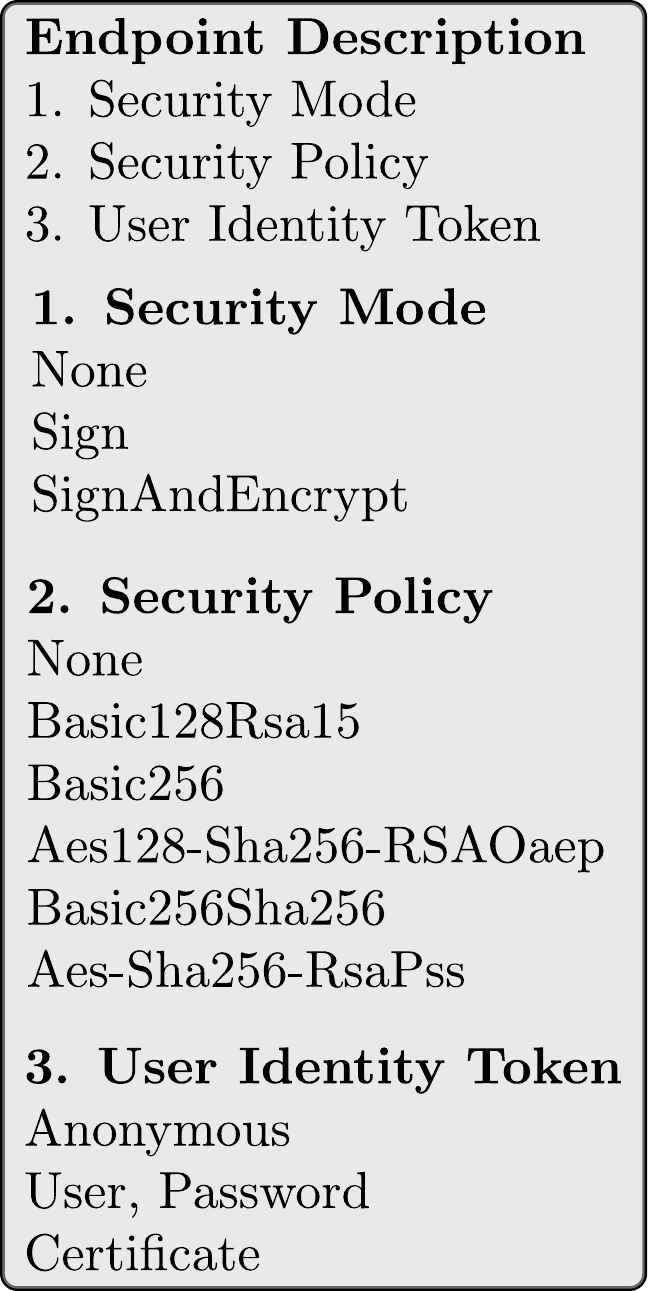}}
    \caption{(a) Overview of the connection establishment procedure. The server provides a list of endpoints. The client selects which endpoint to connect to. If a secure endpoint is chosen, the application authentication through certificates is performed. 
    (b) Options available for the endpoint configuration. The Security Mode, Security Policy and User Identity Token are chosen independently from each other.}
    \label{fig:simple_connAB}
\end{figure}

\opc~\cite{opcfoundation} is an industrial communication protocol developed by the OPC foundation that allows platform-independent and secure communication by design. 
The last specification 1.04 was released in 2018 and is divided in eighteen parts. Part 2 describes the security model of \opc, and Part 4 describes the services that implement security primitives. Two communication strategies are possible: client-server and publisher-subscriber. Both allow messages to be signed to ensure authenticity and encrypted to add confidentiality. \opc does not enforce the use of security, messages can be exchanged without security. Figure~\ref{fig:simple_connAB}(a)(b) reports an example of secure connection establishment in client-server. A server offers several endpoints, each defined by a Security Mode, a Security Policy, and the supported User Identity Token(s). When initiating a connection, the client chooses the endpoint (from an unauthenticated list of endpoints). The Security Mode defines how messages are exchanged to achieve authentication, confidentiality, and integrity. Available Security Modes are None, Sign, SignAndEncrypt. 
Security Policies define the cryptographic primitives used for the specific security mode.
 The UserIdentityToken defines the supported user authentication methods for an endpoint: Anonymous (no user authentication), Username\&Password, Certificate~\cite{opcpart2}.

\Par{Certificate Management} Secure connection establishment (Security Mode not `None') requires \opc applications (clients and servers) to exchange their Application Instance Certificate. Upon receiving a certificate, an application needs to decide whether the received certificate is trustworthy. To this end, each \opc application has a list of certificates that are trusted (Certificate \trust). This list contains self-signed certificates or certificates from Certificate Authorities (CA). In the first case, the certificates are often exchanged manually between applications, but the use of a Certificate Manager with a Global Discovery Server (GDS) is also possible. GDS provides two main functionalities: i) Application discovery: \opc applications use the GDS to find available applications that have previously registered with the GDS. ii)  Certificate management: applications push and pull  certificate to the GDS to update \trust.

\Par{Secure Connection Establishment} To establish a secure connection, the server sends its certificate through the \verb|GetEndpoints| response. If the client trusts the server and selects a secure endpoint, the \verb|OpenSecureChannel| message 
carries the client certificate. Upon trusting the client's certificate on server side the secure channel is established
. The \verb|OpenSecureChannel| request and response use asymmetric encryption and messages contain nonces  used to compute the symmetric signing and encryption keys used in sessions~\cite{opcpart6}.~Due to space constraints further details can be found in Appendix~\ref{app:servicesets}.

\section{\opc Security Assessment Methodology}
\label{sec:methodology}
\opc features a number of cryptographic options to establish secure channels (see Section~\ref{sec:background}). Like in any system, authentication of parties to each other (application authentication in \opc) requires either pre-shared secrets or certificates with public keys. In this work, we argue that in the ICS setting (in which \opc is adopted), no public PKI (e.g.,~with root CAs) is available, so a core issue is how to establish the initial trust between parties. In \opc there are two ways to perform certificate management and distribution: i) Through self-signed certificates, ii) through the GDS \emph{CertificateManager} (see Section~\ref{sec:background}). The focus of our security assessment is the application authentication functionality required for the security in \opc. 

\subsection{System and Attacker Model}
\label{sec:systemandattacker}
\begin{figure}[tb]
\centering
 \subfigure[]{\includegraphics[width=0.6\linewidth]{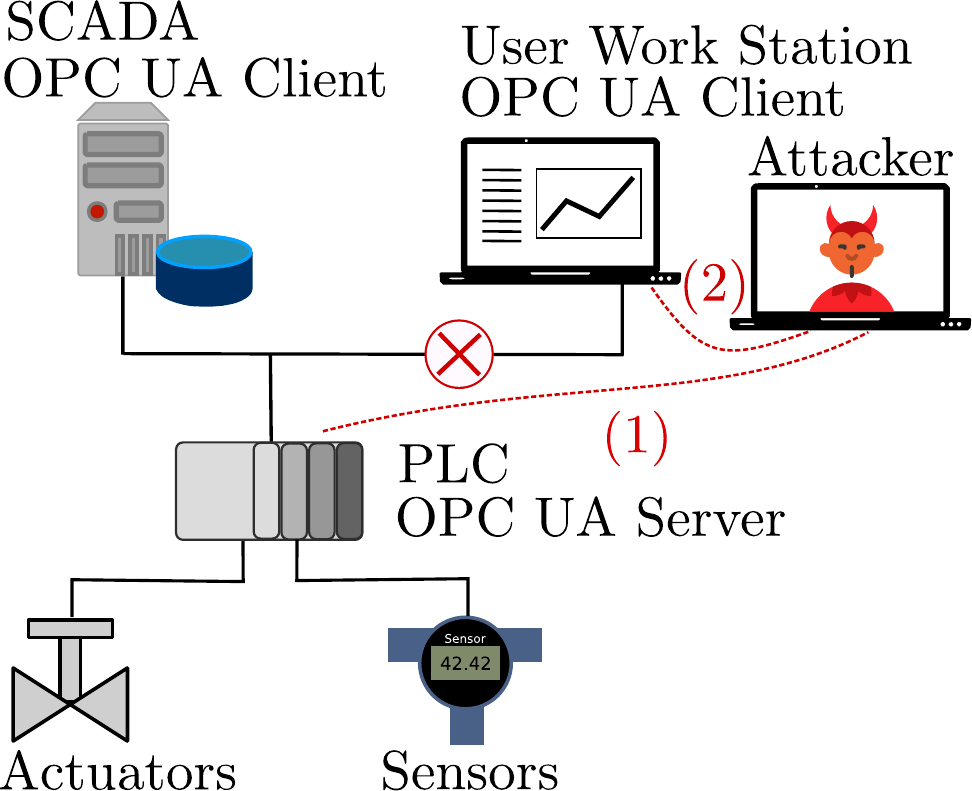}} 
    \caption{ Example \opc network with a server and two clients. An attacker that wants to establish themselves as a PitM has to (1) pose as an \opc client towards the legitimate \opc server and (2) as an \opc server towards a legitimate client.}
    \label{fig:simple_connC}
\end{figure}

We assume a local ICS network with \opc server as depicted in Figure~\ref{fig:simple_connC}. The system operator guarantees security in the \opc system by allowing the server to offer only secure endpoints (i.e., Sign or SignAndEncrypt, see Section~\ref{sec:background}) as suggested by the BSI~\cite{bsi17opcua}. We note that Intrusion Detection Systems are out of scope in our model as we aim to look at security guarantees from the \opc artifacts. We assume the network operator follows the user manual shipped with the deployed ICS hardware/software to configure the \opc server and client security. In this work, we consider three attackers with different goals\footnote{We use the gender-neutral `they' as pronoun in this work}. 

\Par{\rserver} A new device is introduced and now needs to establish secure \opc communications with other devices. An attacker is present in the network and aims to manipulate \opc clients by providing malicious information or stealing of \opc user credentials.  They create a server that offers secure endpoints to establish a secure connection with new clients and make them believe that they are communicating with the actual \opc server in the network.

\Par{\rclient} An attacker aims to connect to the \opc server to eavesdrop or manipulate the information shared between the server and clients. They create a client that attempts to connect to the server although not authorized by the network operator.

\Par{\middlep attack (PitM)} An attacker aims to establish themselves as \middlep between the client and server, intercepting and manipulating all communications between both. This requires achieving \rclient and Server objectives.

\subsection{Research Questions and Challenges}
With our research, we aim to address the following research questions. \textbf{R1.}~\emph{What are practical challenges for the correct use of \opc security features?} \textbf{R2.}~\emph{Are \opc security features correctly implemented by the vendors and products?}  \textbf{R3.}~\emph{What are the consequences of breaking \opc security features?}

While addressing those research questions, we tackle the following research challenges: i)~Partially proprietary products without source code. ii)~available \opc libraries partially documented, iii) unavailability of products (or real deployments) for testing.

\subsection{Proposed Approach}

Our approach focuses on the analysis of security features implemented by \opc compatible artifacts. In particular, we focus on implemented key management functionalities i.e.~i)~Proprietary hardware and software that implements \opc stack, ii)~Open source libraries that implement the \opc stack. 

To address R1, we survey proprietary and open source \opc enabled products.  We investigate practical challenges for correctly configuring secure \opc setups in the analyzed products checking availability of features. 

To address R2, we propose a framework to verify the correctness of implementations of \opc security features in hardware and software products. For proprietary products, we consult user manuals. For libraries, we practically tested the security features deploying \opc clients and server locally. 
We tested the three attacks presented in the Attacker Model, which are feasible due to erroneous or incomplete key management features. For each library, we set up \opc server and client following the instructions.
To address R3, we show that breaking the application authentication configuration will have severe consequences on \opc security. An attacker will be able to obtain plain-text passwords when legitimate clients connect to \rserver or the \middlep system and perform user authentication, to modify the data seen by \opc clients and servers, and to execute function calls on an \opc server that can interfere with the physical process (e.g.~send commands to actuators). 


\section{Framework and Implementation}
\label{sec:implementation}
In this section, we present our framework for the security assessment of \opc artifacts. First, we will describe our framework from a high-level perspective. Then, we provide the details of our framework implementation.

\subsection{Framework} Our framework tests \opc clients and \opc servers against \rserver, \rclient, and \middlep attacks. For \opc servers, the framework identifies vulnerabilities to \rclient Attacks, while for \opc clients the framework identifies vulnerabilities to \rserver Attacks. The framework checks for vulnerabilities generated by incorrect (or incomplete) \trust management. When both \rclient and \rserver vulnerabilities are present in the \opc network, \middlep attacks can be exploited.

\begin{figure}[t]
   \centering
    \includegraphics[width=0.75\linewidth]{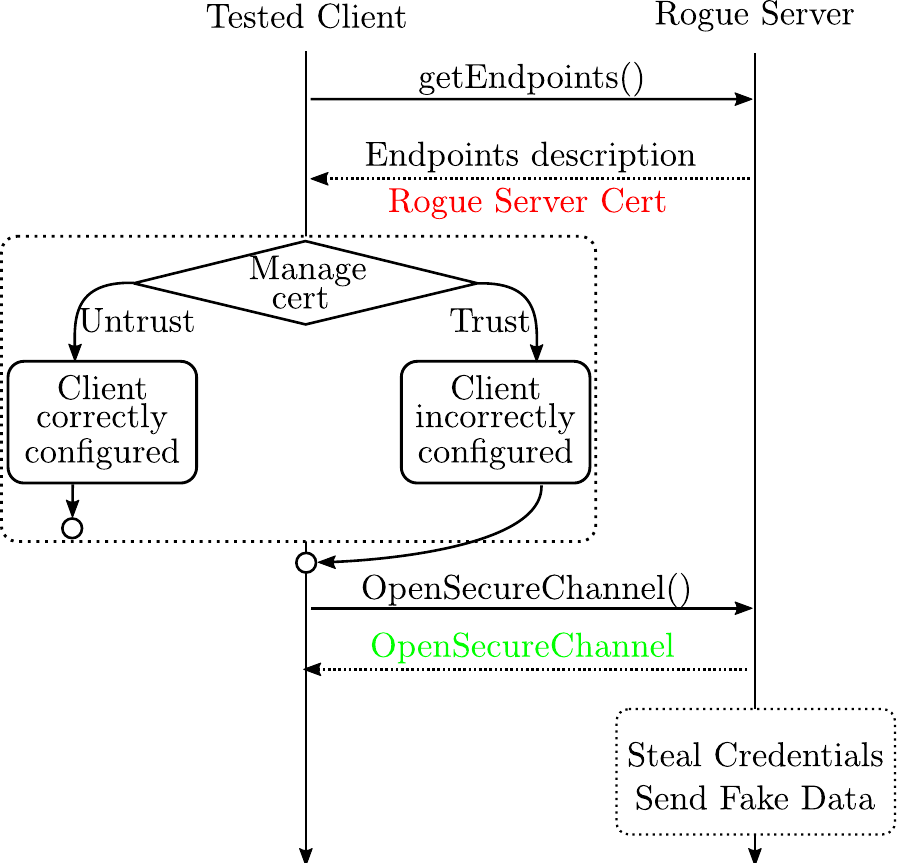}
    \caption{\rserver workflow. The server waits for a new connection. When the Server receives a getEndpoints request it provides a certificate that was not trusted on the client side upon connection (in red). If the client trusts the server certificate and sends an OpenSecureChannel to the server, the \rserver allows the connection (in green). The attacker can, for example, steal credentials and send manipulated data.}
    \label{fig:rogue_server}
\end{figure}

\Par{\rserver}~In Figure~\ref{fig:rogue_server} we report the steps followed to verify the vulnerability to the \rserver. The test employs an \opc \rserver to test \opc clients. The \rserver waits for a \verb|getEndpoints| request from a victim client and replies offering secure endpoints and its self-signed application instance certificate that is not present in the client's \trust. The victim client receives the endpoints and the certificate. The \trust of the \opc victim client does not contain the \rserver's certificate. The victim client should not establish a secure connection with the untrusted \rserver as the root of trust between client and \rserver was not established upon connection. If the victim client is correctly configured, it will not continue the interaction with the \rserver. Otherwise, the victim client trusts the \rserver and instantiates an \verb|OpenSecureChannel| request that the \rserver accepts. 

As a consequence, an attacker can send fake data to the victim client and steal user credentials. User credentials stealing occurs when the \verb|ActivateSession| request with `UserIdentityToken' `user\&password' is instantiated by the Client.  
The request (containing user and password) is encrypted with the keys derived from the \verb|OpenSecureChannel| Nonces. Moreover, the password is encoded in UTF-8 and encrypted with the server public key (i.e.,~the key received from the untrusted \rserver). Despite the different encryption techniques applied to encrypt the password before transmission, the \rserver will decrypt them because the client allows the connection with untrusted parties.

\begin{figure}[t]
    \centering
    \includegraphics[width=0.9\linewidth]{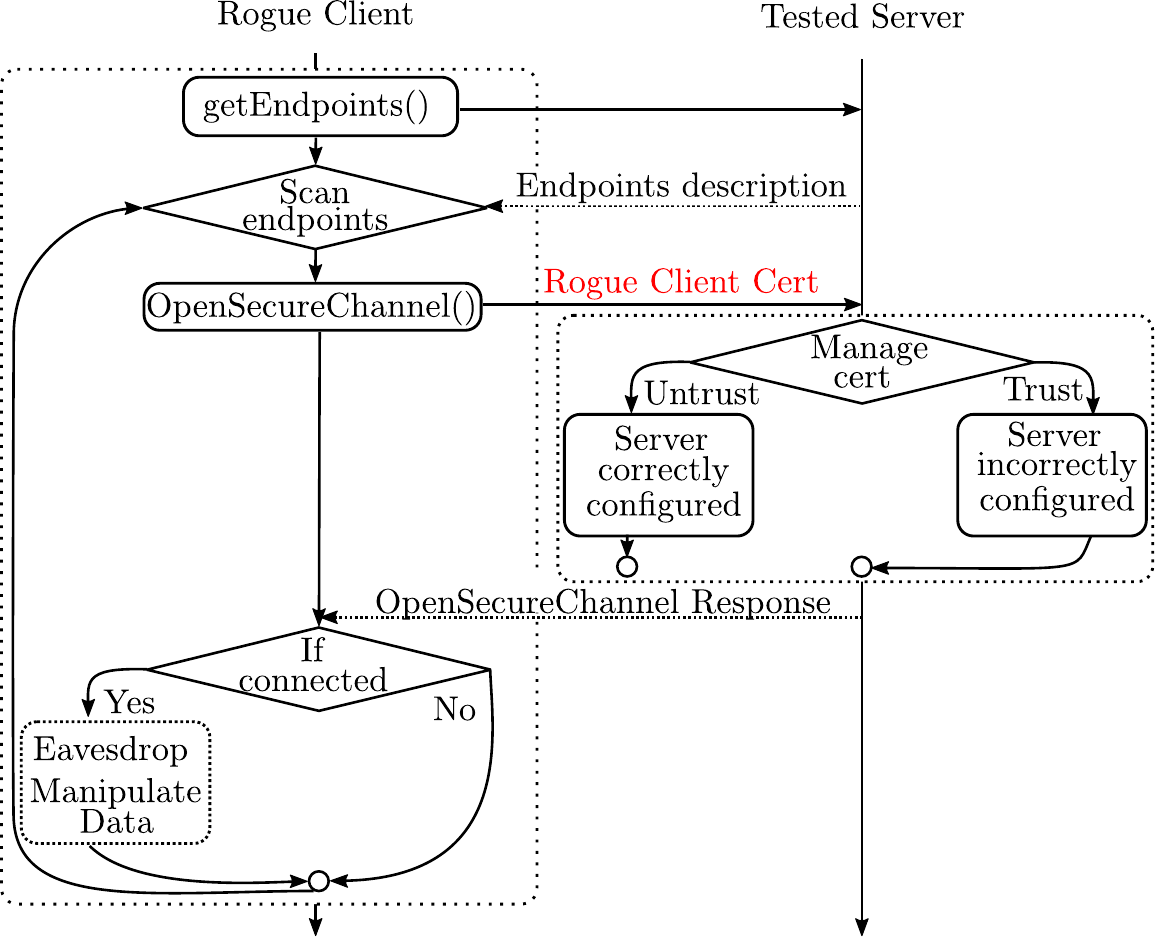}
    \caption{ \rclient workflow. The dashed boxes contain the flowchart representing the application server/client logic. Diamonds represent a decision by the client/server according to the data flow and management. The client lists all the secure endpoints on the server and for each one, it tries to connect providing an arbitrary self-signed certificate (highlighted in red) that was not shared with the server in advance. If the client successfully connects, the server is not correctly managing certificates.}
    
    \label{fig:rogue_client}
\end{figure}

\Par{\rclient}~In Figure~\ref{fig:rogue_client} we report the steps followed by our framework to verify the vulnerability to the \rclient in \opc servers. The test employs an \opc \rclient to the test \opc server implementations. The \rclient scans all the endpoints offered by an \opc server and tries to establish a secure connection to all of them, one by one. The \trust of the \opc server does not contain the \rclient's certificate. Hence, the \rclient should not be entitled to establish a secure connection with the server because the root of trust between client and server was not acknowledged upon connection. If the client succeeds in connecting, the server implementation is managing erroneously the certificates, and it is deviating from the expected behavior prescribed by \opc protocol. The server is then considered vulnerable to \rclient attacks. 

As a consequence of a \rclient attack, an attacker can perform different actions on the server according to the server configuration. The attacker possibilities range from reading values published by the attacked \opc server to writing values and executing commands on the server that influence the physical process.

\begin{figure}
   \centering
    \includegraphics[width=\linewidth]{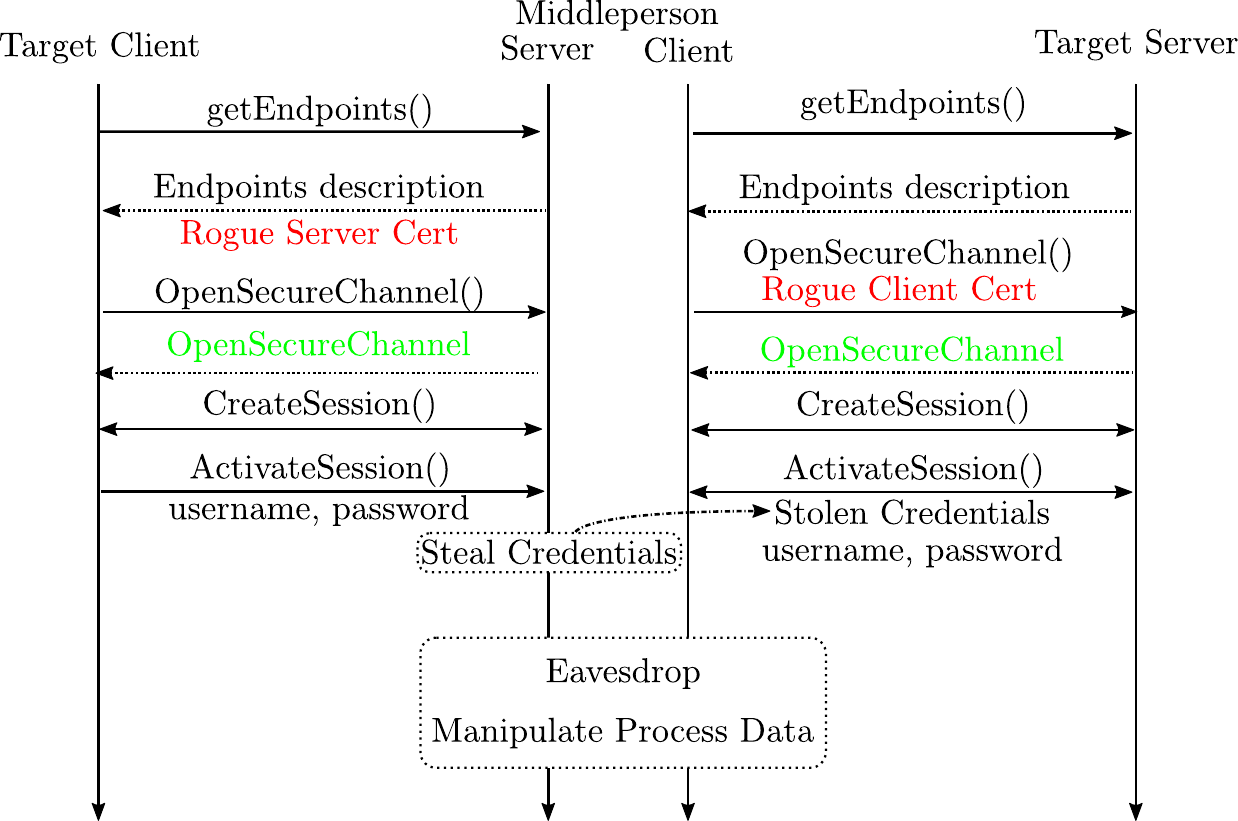}
    \caption{Example of \middlep attack with stealing of session credentials (User Identity Token: User\&Password). The attacker acts both as Rogue Client and Server. The Target Client connects to the \middlep Server. The client is vulnerable to \rserver attack and trusts the server cert without verification and instantiates a OpenSecureChannel() request. The \rserver allows the connection. When the client instantiates a session the \rserver accepts it and decrypts the provided credentials. The \middlep connects to the Target Server (vulnerable to \rclient Attacks) and creates a session providing the stolen credentials}
    \label{fig:middleperson}
\end{figure}

\Par{\middlep}~Figure~\ref{fig:middleperson} reports the steps followed by our framework to verify the vulnerability to the \middlep attack. The test leverages the concepts of \rclient and Rogue Server used at the same time to achieve the \middlep attack. The attacker instantiates a \rserver in the network. The client requests a secure connection with the \rserver (as described in the previous paragraphs). If the attacked server requires user authentication with username and password, the \rserver can request the user identity token from the victim's client and steal the credentials. At this point, the \rclient instantiates a connection with the victim's server by providing the stolen credentials and takes control of the industrial process (as described in the \rclient paragraph). The attacker forwards stealthily (from the process operators) the information received from the victim client to the victim server and vice versa.

Consequences of the \middlep attack comprise the union of the outcomes identified for \rclient and \rserver.

\subsection{Implementation}
We implemented our framework (i.e.,~\rserver, \rclient, and \middlep) using the python opc-ua~\cite{libraries:python-opcua} open source library. The implementation serves as a Proof of concept (POC) of our attacks. We have implemented the POC using the python opc-ua since the library offers support to all the functions required to implement our proposed attacks. Moreover, the functionalities are easy to prototype and deploy. 

Our framework consists four python modules, \verb|framework|, \verb|utils|, \verb|rogueclient|, and \verb|rogueserver| for a total of $571$ lines of code. For each of the three attacks, we report a description of how we realized functionalities through the python modules. 

\Par{Framework module} The \verb|framework| module interacts with the attacker. It offers a command-line interface to select the attack (i.e.,~\rserver, \rclient, or \middlep). 

\Par{\rserver attack} When \rserver attack is selected, the \verb|rogueserver| module performs the following to mount and start the attack. First, the program executes port scanning in the network to identify available \opc servers on port $4840$. The attacker selects a benign server that they want to clone with the \rserver attack. At this point, the sequence of unencrypted and unauthenticated primitives \verb|FindServers()| and \verb|GetEndpoints()| requests are sent to the benign server to retrieve endpoints information, server information, and the server certificate. A \rserver is created with this information. The \rserver has the same name, offers the same endpoints (same security mode, security policy), the same user identity token, and provides a self-signed certificate filled with the same information as the benign server (except fingerprints). When \rserver is configured, port forwarding is enabled in the network to route requests intended to the victim's server to the \rserver. Finally, the server is started and waits for a victim client to connect to it without performing certificate validation. If the connection succeeds, the \rserver starts publishing fake data, and if the victim's client provides user credentials, the \rclient decrypts them.

\Par{\rclient attack} When \rclient attack is selected, the \verb|rogueclient| module performs the following actions to mount and start the attack. First, the attack performs port scanning in the network to identify available \opc servers on port $4840$. The attacker chooses the victim server that they want to connect through the \rclient attack. At this point, the client attempts the \verb|OpenSecureChannel()| request to the victim server endpoints (one by one), providing a self-signed certificate that was not inserted in the victim server \trust upon connection. If the cert is trusted, the attack is successful. Furthermore, if the victim server requires the user identity token 'None': the \rclient instantiates a \verb|ActivateSession()| request and can start reading (or writing) values at server nodes. If a user identity token with username and password is required, the attacker can input them (e.g.,~they retrieved them with a \rserver attack). Finally, the \verb|ActivateSession()| request is sent to the server.

\Par{\middlep attack} When the \middlep attack is selected, our POC realizes the attack depicted in Figure~\ref{fig:middleperson}. The \verb|rogueserver| and \verb|rogueclient| modules are used in parallel (i.e.,~in threads). First, the \rserver is created to capture victim client requests in the network and acquire user credentials. Once the credentials are retrieved, the \rclient is instantiated and connects to the victim's server providing the stolen credentials. The attack is successful if both the victim's client and server do not populate the \trust upon connection.

\section{Assessment Results}
\label{sec:results}
In this section, we present the results of our security assessment of \numproducts \opc products by different vendors (i.e.,~mostly OPC UA servers for PLCs), and 26 systems we build using \numlibraries libraries. We start by reporting the selection criteria of the artifacts considered in this work. Next, we investigate the availability of features of the \opc stack implemented in the considered artifacts. Then we perform a security assessment for \opc products with our framework and look for vulnerabilities to \rclient, \rserver, \middlep attacks.

\subsection{Artifacts Selection Criteria}
For the two categories of \opc  artifacts considered in our work, we used the following selection criteria. Note that not all of the artifacts are certified by the OPC Foundation.

\Par{Proprietary products} We assessed only proprietary products with publicly accessible manuals that provide details on the availability of \opc security features. 

\Par{Libraries} We assessed libraries based on the availability as open-source or free (i.e.,~unlimited unpaid access) versions. For open-source libraries, we consulted Github and selected the libraries based on their popularity (e.g.,~number of stars), and we consulted the list at this repository\footnote{\href{https://github.com/open62541/open62541/wiki/List-of-Open-Source-OPC-UA-Implementations}{open62541: List of Open Source OPC UA Implementations (version at March 10, 2020)}}.
 
\subsection{Adoption of \opc features}
For the considered artifacts we investigate which features of \opc are implemented.

\begin{table*}[t]
\renewcommand{\arraystretch}{1.3}

\begin{center}

    \caption{\opc in proprietary products. \cmark/\xmark~denotes if the product supports/not supports a feature. \tild~denotes that there are problems with feature configuration.}
    \label{tab:vendors}

\begin{tabular}{ll|c|c|c|ccl}
\toprule
Vendor                  & Platform                                        &  OPC Cert.  &     Pub-Sub  &     GDS        &   Security   & \trust       &  Recommended Policy   \\
\midrule                                                                                                                                      
B\&R                    & ADI OPC UA~\cite{vendor:br}                    &     \cmark
&    \xmark    &    \xmark      &    \cmark    &  \cmark      &    Not specified   \\
\hline                                                                                             
Bachmann                & OPC UA Client/Serv.~\cite{vendor:bachmann} &     \xmark      &    \xmark    &    \xmark      &    \tild     &  \tild       &    Not specified    \\
\hline 
Beckoff                 & TC3 OPC UA~\cite{vendor:beckoff}                &     \xmark      &    \xmark    &    \xmark      &    \tild     &  \tild       &   Deprecated protocols  \\
\hline                         
Beijer                  & iX Developer~\cite{vendor:beijer}               &     \xmark      &    \xmark    &    \xmark      &    \xmark    &  \xmark      &    None             \\
\hline                                                                                                                                                                        
Bosch Rexroth           &         ctrlX CORE~\cite{vendor:bosch}          &     \xmark      &    \xmark    &    \xmark      &    \cmark    &  \tild       &    None not supported \\
\hline                                                                                                                               
General Electric        & iFIX~\cite{vendor:generalelectric}              &     \xmark      &    \xmark    &    \cmark  &    \cmark    &  \tild       &    Basic256Sha256     \\
\hline                                                                                                                                                                           
Honeywell               & ControlEdge Builder~\cite{vendor:honeywell}    &      \xmark\textsuperscript{\ding{59}}    &    \xmark    &    \xmark      &    \xmark    &  \xmark      &    None               \\
\hline                                                                                                                                                                       
Lenze                   &  Easy Starter~\cite{vendor:lenze}              &      \xmark     &    \xmark    &    \xmark      &    \tild     &  \tild       &    Deprecated protocols           \\
\hline                                                                                                                                                                       
Mitsubishi              &  MX Configurator-R~\cite{vendor:mitsubishi}    &      \cmark     &    \xmark    &    \xmark      &    \cmark    &  \cmark      &    None               \\
\hline                                                                                                                                                                       
National Instr.    & InsightCM~\cite{vendor:nationalinstruments}    &      \xmark     &    \xmark    &    \xmark      &    \cmark    &  \cmark      &    None               \\
\hline                                                                                                                                                                         
Omron                   & SYSMAC-SE2~\cite{vendor:omron}                 &     \cmark      &    \xmark    &    \xmark      &    \cmark    &  \cmark      &    Not specified     \\
\hline                                                                                                                                                                         
Panasonic               & HMWIN Studio~\cite{vendor:panasonic}           &      \xmark     &    \xmark    &    \cmark  &    \cmark    &  \tild       &    Not specified     \\
\hline    
Rockwell                & Factory talk linx~\cite{vendor:allenbradley}   &      \xmark     &    \xmark    &    \xmark      &    \cmark    &  \cmark      &    Not specified      \\
\hline                                                                                                                    
Schneider               & Control Expert~\cite{vendor:schneider}         &  \cmark
&    \xmark    &    \xmark      &    \cmark    &  \cmark      &     Basic256Sha256       \\
\hline                                                                                                                                                                       
Siemens                 &   STEP 7~\cite{vendor:siemens}                 &     \cmark      &    \xmark    &    \cmark  &    \cmark    &  \tild       &    Not specified      \\           
\hline                                                                                                                                                                        
Weidm\"uller            &  u-create studio~\cite{vendor:weidmuller}      &    \xmark       &    \xmark    &    \xmark      &    \cmark    &  \cmark      &    Basic256Sha256     \\
\hline                                                                                                                                                      
Yokogawa                &  SMARTDAC+~\cite{vendor:yokogawa}              &     \xmark      &    \xmark    &    \xmark      &    \xmark    &  \xmark      &    None               \\    
\hline
\multicolumn{8}{c}{Codesys based platforms} 
\\ \hline
Codesys                 & Codesys V3.5~\cite{vendor:codesys}             &     \xmark      &   \cmark    &    \xmark       &    \cmark    &  \tild       &    Not specified     \\ 
\hline
ABB            &Automation Builder~\cite{vendor:ABB}                     &     \xmark      &    \xmark    &    \xmark      &    \cmark    &  \tild       &    Basic256Sha256     \\
\hline              
Eaton          & XSOFT-CODESYS~\cite{vendor:eaton}                       &     \xmark      &    \xmark    &    \xmark      &    \cmark    &  \tild       &    Not specified       \\
\hline    
Hitachi        & HX Codesys~\cite{vendor:hitachi}                        &     \xmark      &    \xmark    &    \xmark      &    \cmark    &  \tild       &    Not specified      \\
\hline     
Wago           &  e!cockpit~\cite{vendor:wago}                           &     \cmark      &    \xmark    &    \cmark  &    \cmark    &  \tild       &    Not specified      \\

\bottomrule      
\end{tabular}

\end{center}
\textsuperscript{\ding{59}}State of the documentation consulted during the investigation. After a preprint release of this manuscript, the documentation related to the product was updated. Now it supports security and it is certified.
\end{table*}

\Par{Proprietary products} In Table~\ref{tab:vendors} we summarize the adoption of \opc features of the \numproducts proprietary \opc products that are offered by vendors to configure their industrial devices. The table is organized into four parts: Certification by OPC Foundation, support of publish-subscribe, compatibility with Global Discovery Server, security features. Out of \numproducts software packages that were analyzed, four vendors rely on Codesys automation software to implement \opc features. None of the considered products support the publish-subscribe model. This feature, announced in the first semester of 2018, is not yet supported by products. In October 2020, Codesys software released the OPC UA PubSub SL~\cite{vendor:CODESYSPubSub} extension that supports Publish-Subscribe, currently vendors do not integrate it yet. Four vendors mention compatibility with GDS servers, but it is not clear if these vendors also offer their implementation of a GDS server or provide functionality for their products to connect to third-party software.
 
Concerning the security features, out of \numproducts \opc servers, three vendors (Beijer, Honeywell, and  Yokogawa) do not support security features. It means that deploying an \opc network with industrial devices from those brands will always result in an insecure deployment since the only supported security policy is None. Two vendors (Beckoff, and Lenze) support security with deprecated cryptographic primitives, making their applications de facto insecure. For the remaining 16 products that support security features, we have looked at how the user manual guides the customers through the server configuration. Specifically, we observed that two vendors (Mitsubishi and National Instruments) instruct their users to configure security None. Then, three vendors (B\&R, Omron, and Panasonic) discourage the use of None, and seven vendors (Bachmann, Codesys, Hitachi, Rockwell, Siemens, Wago, Eaton) do not give recommendations on the preferred policy. Furthermore, four vendors (ABB, General Electric, Schneider, Wiedm\"uller) recommend a specific Security Policy (Schneider and Weidm\"uller use security mode SignAndEncrypt as default). Finally, one vendor (Bosch Rexroth) does not support security mode None, thereby enforcing authenticated connections. We report in the table if the certificate \trust is supported as required by the \opc specification. As we can see from the table, most vendors support it but, there are several problems in the configuration procedure detailed in user manuals that make deployments vulnerable to the three attacks considered in our manuscript. We will detail the configuration issues in the following subsections.

\begin{table*}[t]
\renewcommand{\arraystretch}{1.3}
\caption{OPC UA Libraries.
\cmark/\xmark~denotes if the product supports/not supports a feature. \tild~denotes that there are problems with feature configuration. Security column reports if the library implements security features. \trust column reports if the library implements application authentication. Demo column reports if demo application supports secure connection}
\label{tab:impl}
\begin{center}
\begin{tabular}{ll|c|c|c|ccc|ccc}
\toprule
                                          &        &       OPC       &    Pub     &          & \multicolumn{3}{c|}{Server}        & \multicolumn{3}{c}{Client}     \\ 
Name                                      & Lang.  & Cert. & Sub & GDS      & Security & \trust  & Demo App.&  Security & \trust & Demo App.\\
\midrule                                                                                                                                                                                
ASNeG~\cite{libraries:asneg}              &C++     &\xmark         & \xmark* & \xmark   & \cmark    &\xmark  &      -        &  \xmark*  &  -     & -        \\
\hline                                                                               
Eclipse Milo~\cite{libraries:eclipsemilo} &  Java  &\xmark         &\xmark   & \xmark   &  \cmark   &\cmark  &      \cmark   &  \cmark   &  \tild & \xmark   \\
\hline                                                                                                                            
Free OpcUA~\cite{libraries:freeopcua}     &  C++   &\xmark         & \xmark  & \xmark   &      -    &    -   &         -     & \xmark    &  -     & -        \\
\hline                                                                               
LibUA~\cite{libraries:libua}              &  C\#   &\xmark         &\xmark   & \xmark   & \cmark    &\xmark  &      \xmark   &  \cmark   &  \xmark& \xmark   \\
\hline                                                                               
node-opcua~\cite{libraries:node-opcua}    &  .js   &\xmark         &\xmark*  & \xmark   &  \cmark   &\cmark  &         \tild &  \cmark   &  \xmark&\xmark    \\
\hline      

opc-ua-client~\cite{libraries:opcuaclient}&  C\#   &\xmark         & \xmark  &    -     &       -   &    -   &            -  &  \cmark   &  \cmark& \tild    \\
\hline   
opcua~\cite{libraries:rustopcua}          & Rust   &\xmark         &\xmark   & \xmark   & \cmark    &\cmark  &        \tild  &  \cmark   & \cmark &\tild     \\
\hline                                                                                                                            
opcua~\cite{libraries:goopcua}            & Golang &\xmark         &  \xmark &  \xmark  &     -     &   -    &          -    &  \cmark   &      - &     -    \\
\hline                                                                                                              
opcua~\cite{libraries:Tsopcua}            &TypeScript&\xmark       &  \xmark &    -     &     -     &     -  &           -   &  \xmark   &  -     & -        \\
\hline                                                                               
opcua4j~\cite{libraries:opcua4j}          &  Java  &\xmark         & \xmark  & \xmark   &  \xmark   & -      &           -   &    -      &  -     &      -   \\
\hline                                                                                                                            
                                                                            
open62541~\cite{libraries:open62541}      &   C    &\tild \textsuperscript{\ding{59}}         &\cmark   &\xmark    &  \cmark   &\cmark  &      \tild   &  \cmark    & \cmark & \tild   \\
\hline                                                                                
OpenScada UA~\cite{libraries:OpenScadaUA} &  C++   &\xmark         &\xmark   &\xmark    &  \cmark   &\xmark  &      \xmark   & \cmark    & \xmark & \xmark  \\
\hline                                                                                
Python-opcua~\cite{libraries:python-opcua}& Python &\xmark         &\xmark   &\xmark    &  \cmark   &\xmark  &      \xmark   &  \cmark   &  \xmark& \xmark  \\
\hline                                                                                 
S2OPC~\cite{libraries:S2OPC}              &C       &\tild\textsuperscript{\ding{59}}          &\cmark   &\xmark    & \cmark    &\cmark  &       \cmark  &  \cmark   & \tild  & \tild   \\
\hline                                                                               
UA.NET~\cite{libraries:UA.NET}            &  C\#   &\cmark         &\xmark   &\cmark    & \cmark    &\cmark  &      \cmark  &  \cmark    &\cmark  & \tild   \\ 
\hline
UAexpert~\cite{libraries:UAexpert}        &   C++  &\cmark         &   \xmark&  \xmark  &      -    &   -    &        -     &  \cmark    & \cmark & \tild   \\ %
\bottomrule
\end{tabular}
\end{center}
\textsuperscript{\ding{59}}Server certified, client not certified. \textsuperscript{*}Denotes that the feature is going to be introduced in the next release
\end{table*}

\Par{Libraries} In Table~\ref{tab:impl} we report the results of our research about the adoption of \opc features that we conducted over \numlibraries libraries to deploy \opc in industrial plants. The table is divided into four parts: support of publish-subscribe, compatibility with Global Discovery Server, server security features, client security features.

Out of \numlibraries libraries: 11 libraries implement server features, and 15 libraries implement client features. Specifically, 10 libraries offer server and client features, 5 offer solely client features, and 1 offers solely server features. Publish-Subscribe is implemented by 2 libraries (open62541, S2OPC), also in this case, this feature is not widely adopted.  At the moment of writing, the OPC Foundation official implementation (UA .NET) does not support this connection mode. GDS is implemented by 1 library (UA .NET). With respect to security features implemented in servers, 10 out of 11 offer security features. Regarding the security features implemented in clients, 12 out of 15 implement security features. Moreover, we have investigated the correctness of security features implementation. Specifically, we looked at the availability of the \trust for certificate verification as described in the \opc standard. Among server implementations, 6 (Eclipse Milo, node-opcua, Rust opcua, open62451, S2OPC, and UA .NET) offer this feature, while the other 5 (ASNeG, LibUA, OpenScada UA, Python-opcua) do not allow the server to verify to which clients they are communicating with. Among the client implementations, 6 (Eclipse Milo, Rust opcua, open62541, UA .NET, opc-ua-client, and UAExpert) offer \trust functionality, while 4 libraries (LibUA, node-opcua, OpenScadaUA, Python-opcua) do not provide the feature to verify the party that they are communicating with. Regarding the remaining 2 implementations (Golang opcua, and S2OPC), we run into issues while verifying their secure connection functionalities that prevented us from testing the availability of \trust features. Finally, we check the security features and verify their correct configuration in the demo applications provided in their repository. 

\subsection{Vulnerability to \rserver}
We tested the vulnerability of 15 Client implementations from libraries, as vendors do not offer \opc client functionalities (apart from Bachmann). We found that all available Client demo applications are configured to connect to an endpoint with security mode None, i.e.,~no security. For 3 libraries  (ASNeG, Free OpcUa, opcua ts) no other mode is possible. The ASNeG library will support security features starting from the next release. The 12 remaining libraries support secure connections. 

For those 12 libraries, we verified their support to \trust for certificate management and tested it with our framework implementation. In 4 libraries (LibUA, node-opcua, OpenScada UA, and python-opcua), the \trust is not supported. 

In 3 libraries, we had issues configuring and running the client application in a secure configuration. In Eclipse Milo, the demo client does not perform certificate validation and the documentation does not provide details about how to enable the \trust on although the feature is present in the source code.  In S2OPC the source code features the \trust management, but we were not able to test it due to errors and missing details for the configuration. Finally, in Golang opcua we were not able to find information related to the \trust neither in the documentation, nor in issues in the repository, nor in the code, hence we assume that this feature is not supported.

In 5 clients, the \trust is supported, but we found two types of insecure behavior that make these clients vulnerable to \rserver attacks: 

\textbf{i) \trust disabled by default.} In 4 libraries (Opcua Rust, UA.Net, and open62541, opc-ua-client) the demo client accepts all server certificates by default. The user can disable this option, then the libraries behave correctly w.r.t. the certificate validation procedure. While this setting is only meant to be used during development and testing, it is an additional hurdle that can lead to a seemingly secure application that is vulnerable to a \rserver attack.

\textbf{ii) Use of Secure Channel primitives to perform certificate exchange.} In one implementation (UAexpert with GUI interface) the instructions guide the user to perform a \verb|GetEndpoints| request to retrieve the Server certificate.  The server replies to the client sending his certificate to the client. The Client prompts the error `BadCertificateUntrusted' since the server certificate fails the security check. At this point, the client is asked to trust the certificate and re-instantiate a secure connection. This behavior is susceptible to \rserver attacks since the certificates are exchanged through an insecure channel. UAexpert offers the option to trust certificates before a connection to server, but this is not the default workflow in the instructions. 

Overall all 12 clients that support security features exhibit vulnerabilities that can be exploited in a \rserver attack. Even libraries that have handled security correctly on the server-side are lacking security features on the client-side, forcing users to lower the security properties of their \opc deployments.

\subsection{Vulnerability to \rclient}

\Par{Proprietary products} Since we do not have access to actual devices, our analysis relies on the official user manuals shipped with products. The results for the \numproducts analyzed products are reported in Table~\ref{tab:vendors}. The vendors Yokogawa~\cite{vendor:yokogawa}, Honeywell~\cite{vendor:honeywell} and Beijer~\cite{vendor:beijer} do not support any security features for \opc: this means that they do not offer secure communications channels to send information to clients and will not be able to perform application authentication. 

For the remaining 19 companies, we have investigated if they use the \trust to enable only certain clients to connect to the server. All the companies implement the \trust for certificate verification, only 7 out of 19 correctly instruct the users to configure it. We found that 12 companies report insecure instructions to perform the certificate exchange necessary to build the \trust that makes them vulnerable to \rclient attack. In particular, we have identified two different issues with the instructions:

\textbf{i) \trust disabled by default.} The instructions by default guide the user that enables security to configure the server to accept all certificates and optionally the user can configure the \trust.  If the server has default settings, an attacker can connect a client to the server providing an arbitrary certificate that is not verified upon trusting it. Siemens and Bachmann's products are affected by this issue.

\textbf{ii) Use of Secure Channel primitives to perform certificate exchange.} The issue resides in the procedure used to exchange certificates. The product affected by this issue leverages the unauthenticated \verb|OpenSecureChannel| request to move the client certificate from client to server (instead of building the \trust before any connection in the network). This behavior can be leveraged by an attacker to install its \rclient certificate on a target server. On the server-side, the certificate is manually or automatically trusted. An operator would need to carefully check the certificate thumbprint to notice that the installed certificate is not authorized. We identified this as a common behavior in the instructions from 10 different vendor products (ABB, Beckoff, Bosch Rexroth, Codesys, Eaton, General Electric, Hitachi, Lenze, Panasonic, and Wago). All 4 products analyzed based on Codesys software propagate this insecure behavior present in the Codesys documentation.  This problem may potentially be threatening more than 400 device manufacturers that rely on this Codesys software. 

\Par{Libraries} We tested the vulnerability to \rclient attack of the 10 server libraries that offer security features. This evaluation is done with our framework implementation as described in Section~\ref{sec:implementation}. To perform our test, we started from demo applications shipped with the libraries and then verified if the library itself has additional capabilities not used in the demo to manage security features. If additional features are available, we add them to the server to test the library. 

All libraries provide a demo server or a tutorial explaining how to set up a simple \opc server. Five demo applications offer secure and insecure endpoints, and five offer exclusively security None. For demo applications where only insecure endpoints are supported, any client can connect to the server via the security mode and policy None (i.e.,~without securely authenticating). Next, we tested the certificate management functionalities offered through \trust (mode Sign or SignAndEncrypt). We found that 6 libraries support the \trust of certificates, and 4 do not support it.  

With our framework \rclient we tested the 6 demo servers provided by the implementations that support the \trust. Our framework tries to establish a connection in mode Sign or SignAndEncrypt where the client provides a self-signed certificate, which was not listed on the server before connecting. According to the \opc standard, such connections should be rejected by the server. In 3 cases the demo server rejects connection from untrusted clients (Eclipse Milo, UA.Net, s2opc). The Eclipse Milo demo server reachable online rejects unknown clients, and users are required to upload their client certificates to the server to appear in the server \trust and connect. Similarly, the demo servers provided by the UA.Net library and the s2opc library reject connection attempts of unknown clients. Moreover, we found that 3 demo servers show the same two types of insecure behaviors identified in the vendor section. 

\textbf{i) \trust disabled by default.} Untrusted connections are allowed (by default) in 2 libraries (node-opcua, open62541),  which are not enforcing the use of the \trust to start the server. Again we found a major flaw in the certificate management. In node-opcua there is a boolean variable 'automaticallyAcceptUnknownCertificate' that is turned to 'true' by default when creating \opc server. This setting is transparent to users that are allowed to build a server without setting this variable explicitly. In the open62541 library, the user can start the server with or without the \trust. The \trust can be selected as an optional parameter to start the server from the command-line interface. When the server is started without the \trust, any incoming certificate is accepted, the program notifies the user is notified of this behavior. 

\textbf{ii) Use of Secure Channel primitives to perform certificate exchange.} The Rust opcua library uses the \trust for certificate validation in demo applications, in our opinion the suggested procedure on certificate exchange is insecure. After a connection attempt, the client certificate is stored in the 'rejected' list of certificates from where an operator should move it to the trusted folder. An attacker can create a certificate similar to the real client certificate that is difficult to tell apart.

Finally, for the remaining 4 demo applications, the missing support to client certificate authentication is caused by missing features in the library altogether. In particular, we found that there are 4 libraries where the implementation of the \trust is missing.  In the library LibUA, a function is prepared for the user with a comment to implement certificate authentication. Users of the ASNeG and the Python-opcua library have pointed out the missing security features in issues on GitHub. For the ASNeG library, this feature is planned for the next release~\cite{issue:ASNeG}. Developers of the Python-opcua library in 2017 acknowledged the issue~\cite{issue:pythonopcua}, but it is not available yet.

\subsection{Vulnerability to \middlep attack}

In Section~\ref{sec:implementation} we explained that the \middlep attack can be performed when there are servers vulnerable to \rclient attacks and clients vulnerable to \rserver present at the same time in the network. Combinations of aforementioned \opc servers and clients that are vulnerable respectively to \rclient and \rserver attacks would make a deployment that is vulnerable to \middlep attacks. In particular, we have identified that 19 servers artifacts out of 29 (65\%) that support security features are vulnerable to the \rclient attack (4 artifacts due to missing \trust features and 15 due to insecure instructions \trust) and 12 clients out of 12 (100\%) that support security are vulnerable to \rserver attack (7 artifacts due to missing \trust features and 5 due to insecure instructions to configure \trust). As we can see there is a non-negligible risk to deploy an insecure \opc network where an attacker can operate as \middlep.

\subsection{Summary of Findings} 

\begin{figure}[t]
    \centering
    \includegraphics[width=7.42cm]{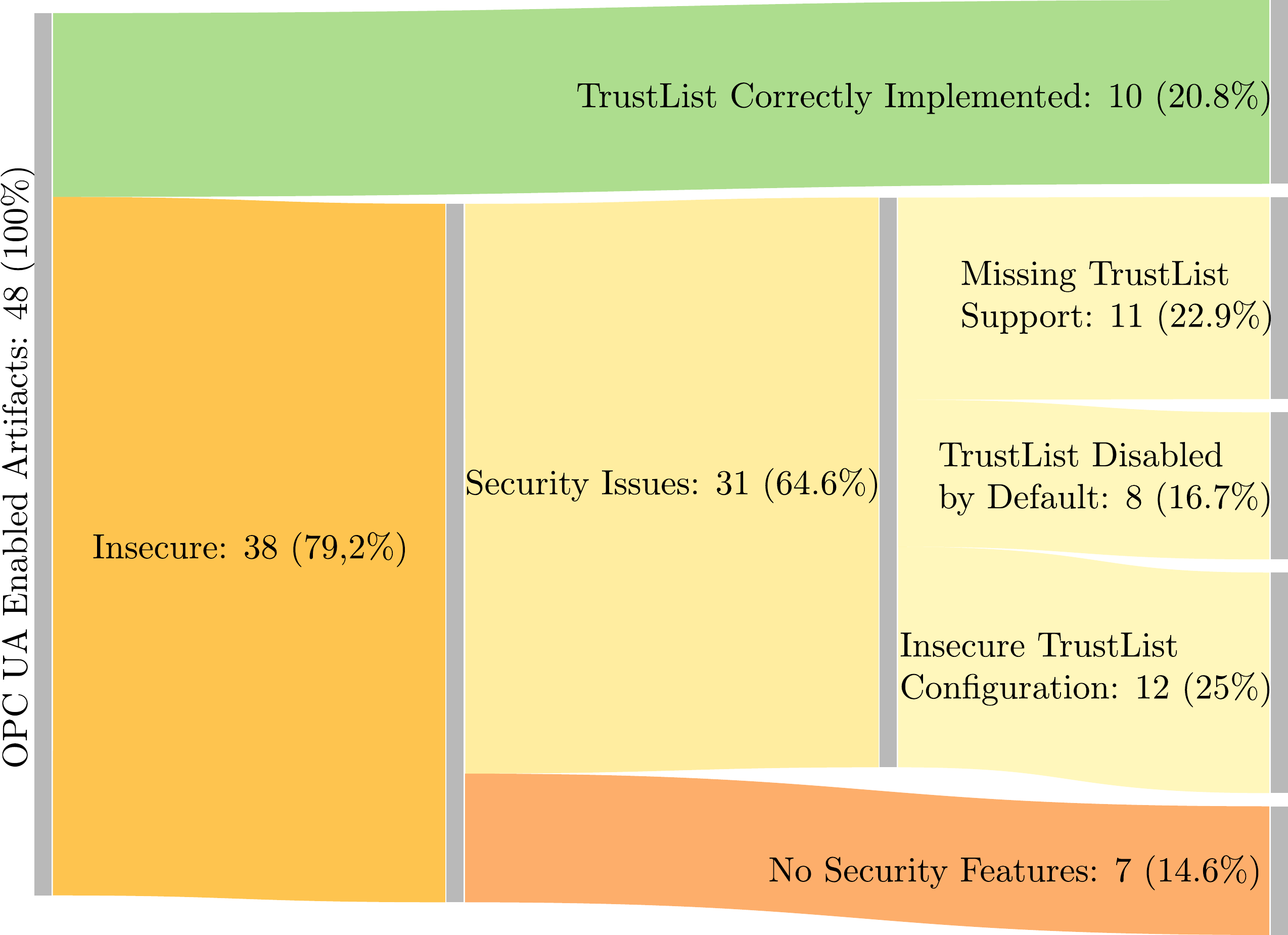}
    \caption{Summary of findings. It shows how the security properties for the 48 \opc enabled artifacts are distributed. For the artifacts that are insecure the chart details the reason of the insecure behavior. The majority of the \opc artifacts present security issues.}
    \label{fig:summary}
\end{figure}

We considered a total of 48 \opc enabled artifacts: \numproducts products from vendors and \numlibraries libraries (of which 11 provide \opc servers, 15 provide \opc clients). Figure~\ref{fig:summary} reports a summary of our findings that we detail in the following. 

Seven artifacts do not support security features at all (14.6\%). Among the 41 remaining artifacts that support security features, we found that 31 artifacts (64.6\% of the 48 \opc artifacts) show issues or errors in the \trust management that enable \rclient, \rserver, and \middlep attacks. The other 10 artifacts (20.8\% out of 48) 
correctly implement the \trust management and instruct users about its configuration, and thus they are not vulnerable to the \rclient, \rserver, and \middlep attacks.
The 31 artifacts that show security issues  with the configuration of the \trust can be classified into the following three categories:
\begin{itemize}

\item \textbf{Missing Support for \trust.} 11 
artifacts do not implement \trust (or do not provide instructions about its configuration) management although they implement \opc security features, that is they offer functionality for signing and encryption but not for the validation of certificates. This makes their deployments always vulnerable to \rclient, \rserver, \middlep attacks.

\item \textbf{\trust disabled by default.}

in 8 artifacts the \trust is disabled by default. This behavior puts \opc deployments at risk as applications will accept any incoming certificate, making them vulnerable to the three considered attacks.

\item \textbf{Certificate exchange through Secure Channel primitives.} In 12 
 artifacts the instructions guide the users to use of unauthenticated Secure Channel primitives to perform certificate exchange. The user is guided to initiate a connection such that the connecting applications sent their certificates to each other. Then the certificates are manually trusted for each device. Since the certificates are sent via an insecure channel an adversary can leverage this behavior to mount an attack. The \opc standard allows this behavior~\cite{opcpart12} assuming that only trained personnel is authorized to trust incoming certificates exchanged through insecure channels. In our opinion, this recommendation is flawed as insecure channels allow simple  manipulation of the certificates before acceptance by the administrator (via \middlep attacks), and humans have been shown to be bad at detecting manipulated certificates~\cite{dechand2016empirical,cherubini2018towards}.
 
\end{itemize}
\Par{Adoption of features} in our analysis of \opc features, we found that the client-server model is available in all tested products while the publisher-subscriber model is supported by 3 artifacts. Moreover, the features offered by the GDS to manage certificates are supported by 5 artifacts. Security features are widely adopted, but they require the correct distribution of certificates in order to deploy a secure network. As we found, this is not always the case in \opc products as many of them do not implement the \trust for certificates or do have issues in the configuration procedure.

\section{Countermeasures}
\label{sec:contermeasures}

Our work showed general missing support and incompleteness of \opc security features in artifacts that make the certificate management often not possible or not required by default. Moreover we discovered some insecure behaviors allowed by the \opc standard. In this section, we discuss countermeasures to prevent vulnerabilities in \opc deployments and achieve an initial secure key distribution.

\subsection{Certificate exchange via insecure channels} 

On the Internet, initial distribution of public keys is commonly solved by shipping devices (or OS) with certificates of a set of core root certificate authorities (CAs). Servers (identified by unique DNS names) then provide certificates authenticated directly (or indirectly) by those root CAs. Such a solution is not possible for ICS networks, as they are air-gapped and do not provide (externally verifiable) unique addressing. \opc Global Discovery Servers with Certificate Manager would be an alternative to manual certificate exchange, but as we showed in Section~\ref{sec:results} this feature is not widely implemented and in any case the GDS certificates will not be shipped together with devices newly introduced into the system. That implies that bootstrapping the security in an \opc system critically relies on manual pre-distribution of certificates even when GDS is deployed.

Currently the \opc standard do not provide effective solutions to overcome this challenge, instead (as explained in Section~\ref{sec:results}) it guides the users to exchange certificates through insecure channels. Given the technological challenges posed by ICS networks, we suggest to i) update the standard (and manuals) to instruct users to rely on out-of-band secure channels for initial certificate distribution, ii) enforce \trust population before use of \opc secure channels and not through unsecured primitives as found commonly in the consulted user manuals and the standard itself. 

Out-of-band certificate exchange can be achieved with different solutions e.g., through secure protocols such as SSH, or through physical solutions such as USB sticks and QR codes~\cite{Marktscheffel16QR}. Each of those possible solutions have different trade-offs in terms of usability and security but in any case offer better guarantees than exchanging certificates through unsecured primitives. Of course, exchanging \opc certificates trough SSH requires that the communicating devices have already shared SSH secrets. USB sticks are a widely supported solution to distribute certificates in the industrial environment, as a drawback the users are required to physically move around the plant in order to distribute the certificates to the devices that are expected to communicate with each other. Finally, QR codes represent another potentially easy to use and deploy solution to distribute certificates in the industrial environment~\cite{Marktscheffel16QR}. In particular, industrial devices could be shipped with a private public key pair, the public key can be printed on a QR code sticker ad physically applied to the device. During the installation process the operator scans the QR code and installs the certificate in the \trust of the authorized devices in the industrial environment. As a drawback of this method, an attacker could deploy a malicious device in the industrial plant while performing a supply chain attack and replace the QR code sticker with his own certificate on legitimate hardware.

\subsection{Missing support for \trust} End users should only use products and libraries that implement the \trust management as described in the \opc standard. In particular, products certified by the OPC can be expected to support this feature. Artifacts that do not provide this feature should instead implement it in order to allow communicating parties to trust each other upon connection.

\subsection{\trust disabled by default} End users should enable the \trust if disabled upon \opc network configuration. Vendors are recommend enabling the \trust by default, and thus making it mandatory to populate the \trust upon the creation of secure channels. Enabling the \trust functionality will require the user to preform additional steps to set up a new device on the \opc network but those steps are of crucial importance for the security of the ICS.

\section{Related Work}
\label{sec:relatedwork}

\subsection{Security of open source \opc libraries} 
Muhlbauer et al.~\cite{muhlbauer2020open} study the security of four popular open-source libraries (UA .Net Standard, open62541, node-opcua, Python-opcua) which we also inspect in this paper. The analysis focuses on five aspects: dependencies, timeouts, supported Security Policies, message processing, and randomness. The authors identify two vulnerabilities in the implementations: missing upper time limits in Python-opcua making it vulnerable to DoS attacks and missing packet type checks in node-opcua. The issues related to the \trust configuration of the libraries investigated our work were not identified. 

Neu et al.~\cite{neu2019simulating} and Polge et al.~\cite{polge2019assessing} showed the feasibility of DoS attacks by a \rclient. They suggest network traffic anomaly detection as a countermeasure. Polge et al.~\cite{polge2019assessing} implemented a \middlep attack using Eclipse Milo. The attacked \opc applications are using Security Policy None or Aes128-Sha256-RsaOaep. They report that if username and password are sent as part of the ActivateSessionRequest message the password is encrypted even in Security Mode None and can therefore not be recovered (in contrast to our findings). The encryption of the password in Mode None is optional in the Specification~\cite{opcpart4}, this design choice was taken by~\cite{issue:milo}. Encrypting credentials with untrusted certificates is susceptible to \rserver attacks. Based on our findings, a \middlep attack that recovers plain-text credentials is possible. Therefore, a \middlep attacker can provide their own certificate, and the user credentials are encrypted using the public key associated with the attacker's certificate.

\subsection{Secure exchange of certificates}
The security of the \opc protocol relies on certificates that authenticate each application. Prior publications proposed techniques to establish the initial root of trust.
In \cite{karthikeyan2018pki}, a PKI is implemented that offers functionality to sign certificates or verify certificates using the Online Certificate Status Protocol.
Meier et al.~\cite{meier2020portable} propose to connect a new \opc application to a physical device with certificate manager functionality to install a certificate and the \trust before connecting the application to the network.

\subsection{Usability issues leading to insecure systems} Usability of security features is an active research topic that relates to the usability of \opc security features. Several studies point out that the integration of security features into programs and systems has to be facilitated~\cite{green2016developers}. For example, Krombholz et al.~\cite{krombholz2017have} conducted a usability study on the configuration of HTTPS. In the study 28 system administrators were asked to configure a web server with HTTPS. They found that participants struggled to find good resources to learn the process, a large number of configuration options were difficult to comprehend and the default configuration only offered weak security. Additionally, the security benefits which a protocol such as HTTPS offers are misunderstood or underestimated. Based on misconceptions some administrators decide against using secure options~\cite{krombholz2019if, fahl2014eve}. 
Acar et al. \cite{acar2017comparing} compare the usability of five Python cryptographic libraries. They find that APIs which offer fewer options lead to better security results. In addition, the documentation of the library and the availability of example code had a stronger influence on the successful completion of the task than the experience of the participant.

 
\section{Conclusions}
\label{sec:conclusion}
In this work, we have systematically investigated practical challenges faced to use \opc securely. To this end, we introduced a security assessment methodology---our assessment considers three attacks that can target \opc deployments, \rserver, \rclient, and \middlep attacks. 
We systematically address three research questions: \textbf{R1.}~What are practical challenges for the correct use of \opc security features? \textbf{R2.}~Are \opc security features correctly implemented by the vendors and products?  \textbf{R3.} What are the consequences of breaking \opc security features?

To address \textbf{R1}, we conducted the first systematical survey of 48 \opc artifacts provided by vendors or open-source. Our survey investigates the availability of \opc security, publish-subscribe, and Global Discovery Server functionalities. We showed that publish-subscribe and Global Discovery Server are not widely adopted. Furthermore, we showed that 7 \opc artifacts do not support security features of the protocol.

To address \textbf{R2}, we proposed a framework to investigate the identified security issues. With our framework, we show that the identified issues make \opc artifacts vulnerable to the considered attacks. We analyzed 48 \opc artifacts, 41 of those artifacts support security features. We found that 31 out 41 artifacts present three recurring pitfalls in the \trust management to establish the initial root of trust.
To address \textbf{R3}, we designed, implemented, and demonstrated three types of attacks. The attacks allow the attacker to steal user
credentials exchanged between victims, eavesdrop on process information, manipulate the physical process through sensor values and actuator commands, and prevent the detection of anomalies.

Our novel findings demonstrate major security flaws in \opc artifacts that threaten the \opc security guarantees. Those results imply that there are a significant number of OPC UA deployments in industry that either do not provide full security guarantees, or rely on the absence of the attacker at configuration time for new devices to bootstrap their authentication process. 
We believe that our proposed systematic approach is useful to increase the awareness in users, developers and companies about the threats that can be produced by the three identified pitfalls in the initial key establishment and \trust configuration. As complementary result, our POC implementation can be used as a tool to probe for erroneous configurations and improve security in \opc deployments.

\bibliographystyle{ACM-Reference-Format}
\bibliography{bibliography}

\appendix
\section{Service Sets}
\label{app:servicesets}

\opc offers its functionalities through 10 service sets. Each service consists of a Service request and Service response, identifiable via RequestHeader and ResponseHeader.
The ten service sets (and therefore the functionalities) offered by the protocol are Discovery, SecureChannel, Session, NodeManagement, View, Query, Attribute, Method, MonitoredItem, and Subscription. Each service set specifies a different functionality as defined in Part 4 of the \opc protocol~\cite{opcpart4}. For example, a client can read and write values related to the processes on \opc Nodes in the server through the Attribute service set. It is also possible to provide the functionality for the clients to add nodes. Moreover, clients can call methods (Method service set) defined for Object Nodes. Through these functionalities, a client can directly influence the plant operations. The \opc server authorizes clients to use its services via user authentication. Session service set provides the user authentication functionalities. User authentication occurs during session activation, after the initial application authentication.  

\end{document}